\documentclass[letterpaper, 10 pt, conference]{ieeeconf}  

\IEEEoverridecommandlockouts                              

\usepackage{graphicx}
\usepackage{xcolor}
\usepackage{cite}

\title{\LARGE \bf
Obstacle avoidance for blind people using a 3D camera and a haptic feedback sleeve*
}

\author{Manuel Zahn and Armaghan Ahmad Khan
\thanks{*This work was supported by FRAMOS GmbH, Germany}
\thanks{Manuel Zahn and Armaghan Ahmad Khan are with the Center for Digital Technology and Management, Technical University of Munich, 80333 Munich, Germany
        {\tt\small manuel.zahn@tum.de, armaghan.khan@tum.de}}%
}

\begin{document}
\graphicspath{ {images/} }
\maketitle
\thispagestyle{empty}
\pagestyle{empty}

\begin{abstract}

Navigation and obstacle avoidance are some of the hardest tasks for the visually impaired. Recent research projects have proposed technological solutions to tackle this problem. So far most systems fail to provide multidimensional feedback while working under various lighting conditions. We present a novel obstacle avoidance system by combining a 3D camera with a haptic feedback sleeve. Our system uses the distance information of the camera and maps it onto a 2D vibration array on the forearm. In our functionality tests of the haptic feedback sleeve, users were able to correctly identify and localize 98,6\% of single motor vibration patterns and 70\% of multi-directional and multi-motor vibration patterns. The combined obstacle avoidance system was evaluated on a testing route in the dark, simulating a navigation task. All users were able to complete the task and showed performance improvement over multiple runs. The system is independent of lighting conditions and can be used indoors and outdoors. Therefore, the obstacle avoidance system demonstrates a promising approach towards using technology to enable more independence for the visually impaired.

\end{abstract}

\section{Introduction}
Even in the present era, visually impaired people face a constant challenge of navigation. The most common tool available to them is the cane. Although the cane allows good detection of objects in the user's immediate vicinity, it lacks the ability to detect obstacles further away. It also remains incapable of detecting obstacles above waist height \cite{b10}, which have no or minimal ground foundations (e.g. bar tables etc.) \cite{b11}. An additional drawback is a sub-optimal performance for elevated surfaces like stairs and the struggle under varying weather conditions. Snow, for example, can cover the textured pavements made for the visually impaired during winter \cite{b12}. All this means that visually impaired people still require external assistance, which limits their independence. This shows the need for a system to enable obstacle avoidance and navigation.

Over the past few decades, many Electronic Travel Aids (ETA) have been developed with the aim of filling this gap. The solutions can broadly be categorized into two major groups based on the medium of feedback i.e. acoustic or haptic. Acoustic feedback-based approaches include, among others, a single camera, open-source solution, which applies image-to-sound rendering technology \cite{b2} and a stereoscopic camera approach with acoustic transducers \cite{b3}. But work by M. J. Proulx et al. demonstrated that systems like these require long times of training to adapt to the device \cite{b4}. Moreover, many blind people highly rely on their sense of hearing for the perception of their surroundings and even navigate by using human echolocation \cite{b16}. Therefore, audio-based feedback through headphones poses the risk of interfering with their hearing-based abilities and needs \cite{b15}. 

Haptic feedback resolves most of the drawbacks caused by acoustic feedback. Early work goes back to 1969 when Bach-y-Rita et al. proposed a vision substitution system using tactile projection \cite{b20}. Since then the technology has seen tremendous advancement as shown by Bloomfield et al. in their work on a wearable vibrotactile array \cite{b21}. Choi et al. provide a holistic overview of vibrotactile solutions from the past decade \cite{b17}. Recently more sophisticated sensors using distance or depth information, have enabled the development of even more reliable blind navigation tools using haptic feedback. Many methods are based on low-cost ultrasonic sensors. Examples include an implementation of an ultrasonic obstacle detection for indoor navigation \cite{b6} and a prototype using three sonar sensors and two vibrators \cite{b5}. The major disadvantage of ultrasonic sensors is the low resolution since they can only tell if an object is within the view field or not. This does not help much in identifying the size or shape of objects and even makes the object localization slow. Therefore other approaches leverage Microsoft's Kinect sensor, which provides a full depth image. S. Mann et al. presented a Kinect on a helmet with 1-D vibrotactile elements around the head to create an assistant for blind navigation based on the depth information \cite{b7}. N. Kanwal et al. built upon that approach by using both the depth sensor and the built-in camera \cite{b8}. They applied corner detection to the camera image to perform obstacle detection and then estimated the distance using the depth map of the same scene. \cite{b9} presented a guidance system using a chest-worn Kinect and sonication to convey obstacle information to the user via headphones. 

Most of the systems and approaches presented suffer the same problem i.e. the tradeoff between the size of the device and the quality of obstacle detection. Smaller lightweight systems like simple cameras or ultrasonic sensors lack the resolution to provide sufficient information. On the other hand, Kinect based systems are too conspicuous for everyday usage, hence, risking user stigmatization. 

\section{Our solution}

The system we are proposing uses a 3D depth camera (mounted on 3D printed glasses) to gather the information about obstacles and a haptic feedback sleeve to display/translate the information to the user. In the following sections, the 3D camera glasses and the haptic sleeve will be explained. Then the mapping algorithm will be introduced, and finally, performance test results will be discussed.

\subsection{3D-Camera Glasses}
\begin{figure}[htbp]
\centerline{\includegraphics{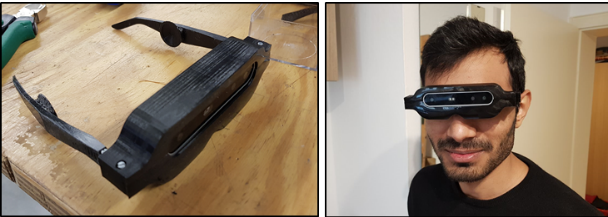}}
\caption{The 3D printed prototype (left) and the prototype in usage (right).}
\label{fig:glasses2}
\end{figure}

The 3D camera glasses are based on an Intel RealSense D415 camera. The camera uses an active IR stereo technology, producing a resolution of 1280x720 and up to 90 fps (for the purpose of this project it was operated at 480p and 6fps to maximize computational performance). It is also low-cost, lightweight and significantly smaller than a Kinect, enabling the development of normally worn glasses. A fitting housing was designed, including spaces (for future work) to add a battery pack, bone conducting speakers and a computing unit. The prototype was 3D printed using fused deposition modeling (FDM) technology and can be seen in figure \ref{fig:glasses2}.

\subsection{Haptic Feedback Sleeve}
\begin{figure}[htbp]
\centerline{\includegraphics[width=0.45\textwidth]{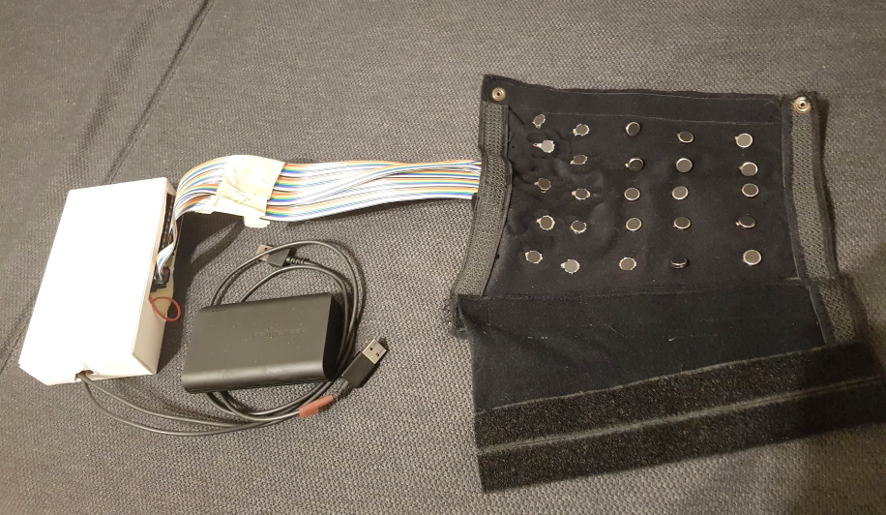}}
\caption{The inside of the prototype.}
\label{fig:sleeve1}
\end{figure}

For transmitting the gathered information to the user, a haptic feedback sleeve was designed. It consists of a 2D array of 25 vibration motors sewed onto a stretchable fabric to allow easy adjustment to varying forearm sizes (figure \ref{fig:sleeve1}). Each motor can be controlled individually using PNP transistors and an Adafruit 24-Channel TLC5947 driver (used to extend the Pulse-Width Modulation (PWM) ports of an Adafruit Feather 32u4 Microprocessor). The slim design allows the user to even wear clothing on top of the sleeve, making it less conspicuous, hence reducing user stigmatization when worn in public. The system is worn on the forearm because humans distinguish vibrotactile feedback better when it occurs close to body landmarks/joints \cite{b14}.

\subsection{The Combined System}
\begin{figure}[htbp]
\centerline{\includegraphics[width=0.4\textwidth]{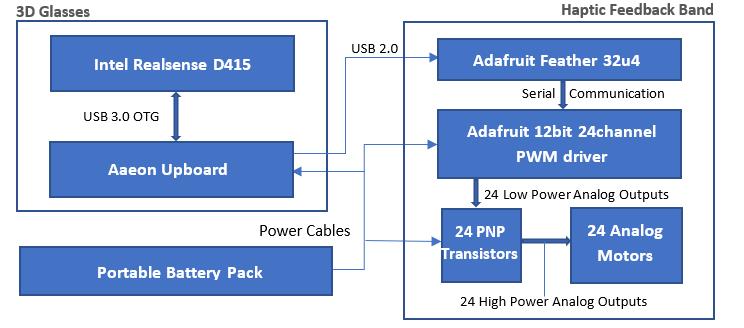}}
\caption{Schematics of the functioning system}
\label{fig:system1}
\end{figure}

To create an obstacle avoidance system, both components are combined. For that purpose, an AAEON Up Board \cite{b18} is used. It runs Linux and serves as a processing unit for downsampling the depth image generated by the 3D camera. The Up Board is also used to transmit the control signal for the motors to the Microprocessor of the haptic feedback sleeve. The complete system is powered by a 5V, 5A battery pack, making the prototype lightweight and portable. A schematic sketch of the system design can be seen in figure \ref{fig:system1}.

\subsection{The Mapping Algorithm}
\begin{figure}[htbp]
\centerline{\includegraphics[width=0.4\textwidth]{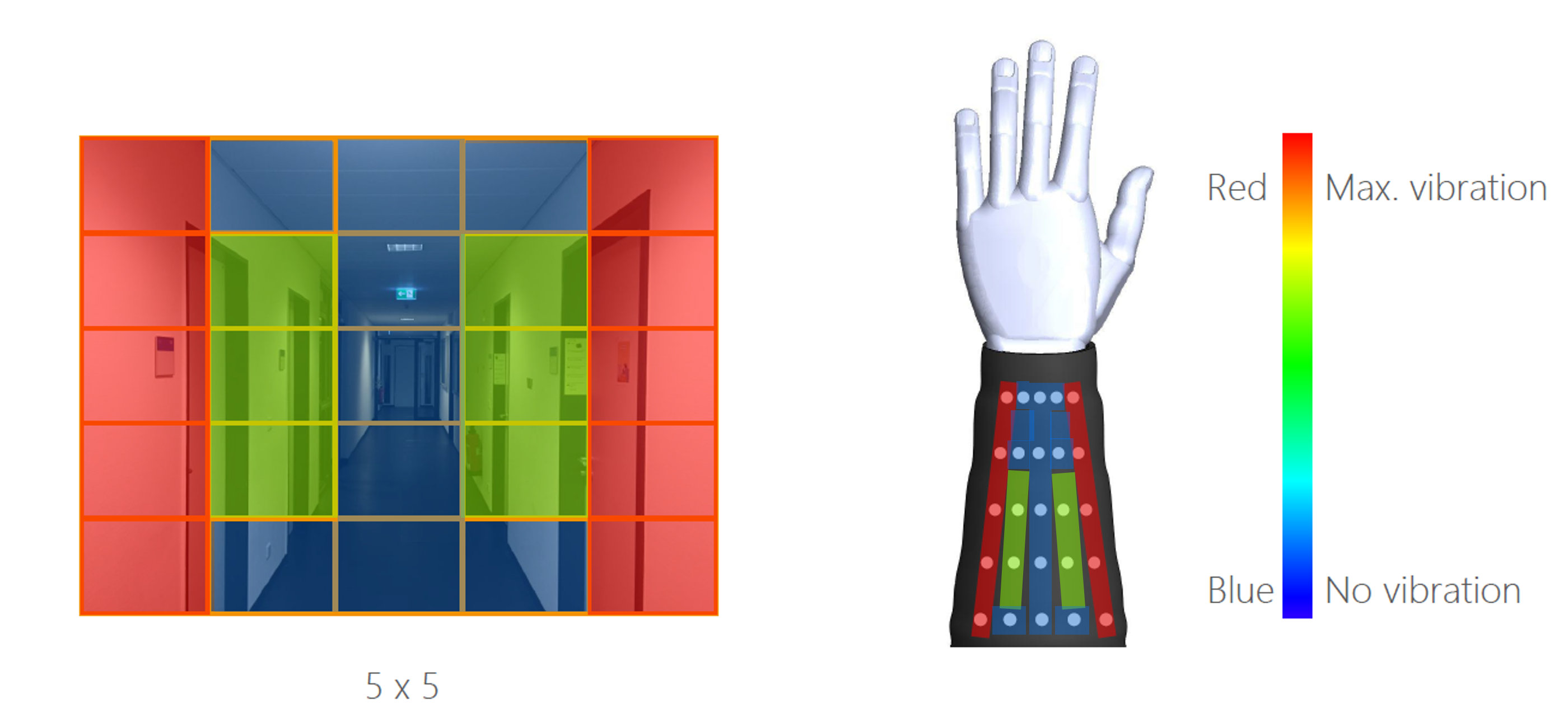}}
\caption{The results of the mapping algorithm for a hallway.}
\label{fig:mapping1}
\end{figure}

Since the RealSense D415 has a resolution of up to 1280x720 pixels, downsampling is necessary to map the information to the haptic feedback sleeve. The input depth image is down-sampled into a 5x5 array, where each array element represents a single vibration motor. These depth values are then used to control the strength of vibration by varying the PWM voltage for the respective motor. A narrow hallway will, thus, generates strong vibrations on each side of the sleeve and the intensity of these vibrations reduces when moving to the center of the array (figure \ref{fig:mapping1}). Users can detect obstacles based on the structural feeling of vibrations. If a user walks towards an obstacle, the vibration intensity of the respective motors gradually increases. Additionally since the arm is only used to display the information of the camera, the user can now walk and perform tasks that require two free hands at the same time. 

\subsection{Operation Mode and Lighting}
To ensure adaptability to environmental variation, it is important to have feedback that can adjust dynamically. For this purpose, two different modes of operation are proposed. In the indoor mode, vibrational feedback is restricted to a certain distance (a distance of 3m is proposed). The vibrational range is then covered within this distance which produces relatively large step sizes for a small change in obstacle distance. In the outdoor mode only distances greater than two meters, out of the range of a cane are mapped. This allows the system to be a good add-on to the existing solutions, by giving the blind user a way to identify distant objects.

\begin{figure}[htbp]
\centerline{\includegraphics[width=0.4\textwidth]{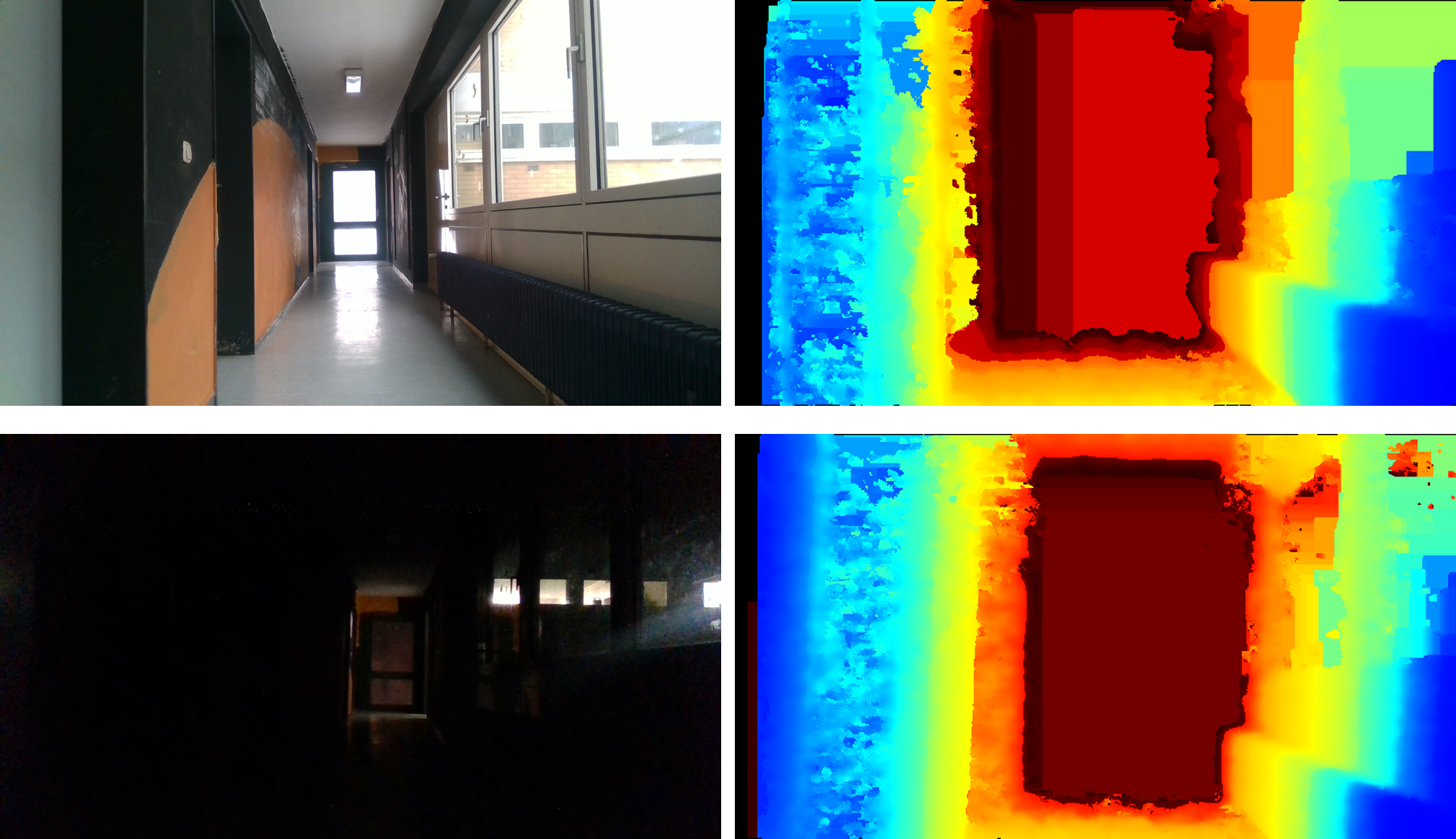}}
\caption{Depth maps for pictures in light (top) and dark (bottom)}
\label{fig:daynight}
\end{figure}

One of the advantages of using a Realsense camera is its IR module. It allows the perception of depth even in complete darkness. As exhibited in the photos in figure \ref{fig:daynight}, the depth results from the day and night imagery of the same scene are almost the same. This feature of the camera makes the solution robust to variation in the lighting conditions.

\section{Method}
To evaluate the performance of the new obstacle avoidance system, two tests were conducted. The goal was to confirm the ability to display information with a vibration sleeve and prove the usability of the combined system as an obstacle avoidance system. Also, the user adoption time for the new system should be tested.

\subsection{General Pattern Tests}
\begin{figure}[htbp]
\centerline{\includegraphics[width=0.3\textwidth]{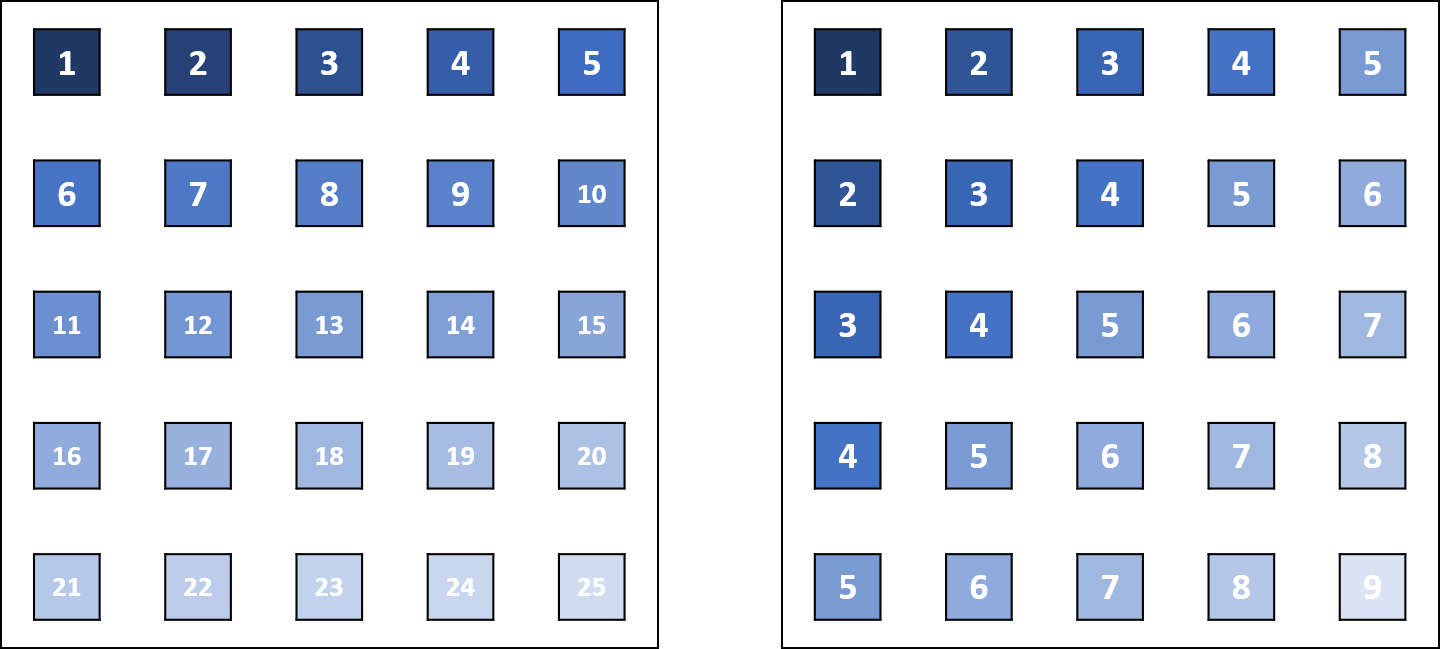}}
\caption{Two example patterns of an iterating single motor vibration on the horizontal axis (left) and lower multiple motor vibration on the horizontal and vertical axis (right)}
\label{fig:patterns8}
\end{figure}

\begin{figure}[htbp]
\centerline{\includegraphics[width=0.45\textwidth]{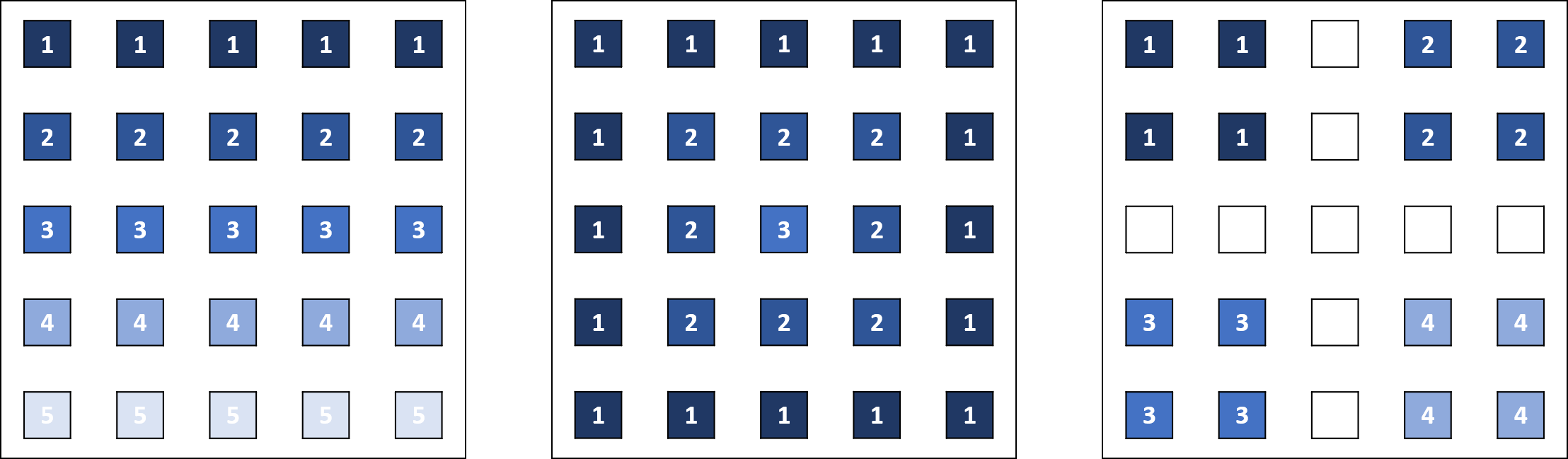}}
\caption{Three example patterns of an iterating multi-motor vibration on the vertical axis (left) a iterating higher multiple motor vibration on the vertical and horizontal axis (middle) and a multiple motor vibration localization task (right)}
\label{fig:patterns4}
\end{figure}

A vibration pattern test was performed to analyze the functionality of the haptic feedback sleeve. The goal was to test if users can localize and tell apart vibration motors on the sleeve. Eleven vibrational patterns were designed and tested on eight users. The experiment was done in three iterations with a pause of one week in between iterations, to identify user adaptation over time. The patterns included single and multi-axis vibrations to test the usability on simple and more complex patterns. Furthermore, single, lower multiple (2-5) and higher multiple ($>$ 5) numbers of motors vibrating at the same time, were used to analyze the localization of vibrations. The user had to pick from one of the three options. An Example of a test pattern can be seen in figure \ref{fig:patterns8}. The left pattern shows a single motor vibration on one axis. The vibration of a single motor moves along the rows of the vibration band, starting in the top-left corner (1) and ending in the bottom-right corner (25), with each motor vibrating for 500ms. The blue color gradient visualizes the moving direction of the vibration. The right pattern of figure \ref{fig:patterns8} shows a multi-axial vibration pattern with a lower multiple of motors vibrating at the same time. The vibration starts again with a single motor at the top-left corner (1) and then continues diagonal towards the bottom-right corner (9), with a maximum of 5 motors vibrating at the same time. Further examples can be seen in figure \ref{fig:patterns4}. All user responses during the tests were analyzed against two criteria to classify them as correct, partially correct and wrong. The criteria being: \\
\begin{itemize}
\item[1)] Correctly identify the pattern based on the moving direction of the motor vibration over time.
\item[2)] Correctly identify the number of motors (single, lower multiple, higher multiple) vibrating at the same time for each time step of a pattern.\\
\end{itemize}

\subsection{Testing Route}
\begin{figure}[htbp]
\centerline{\includegraphics[width=0.35\textwidth]{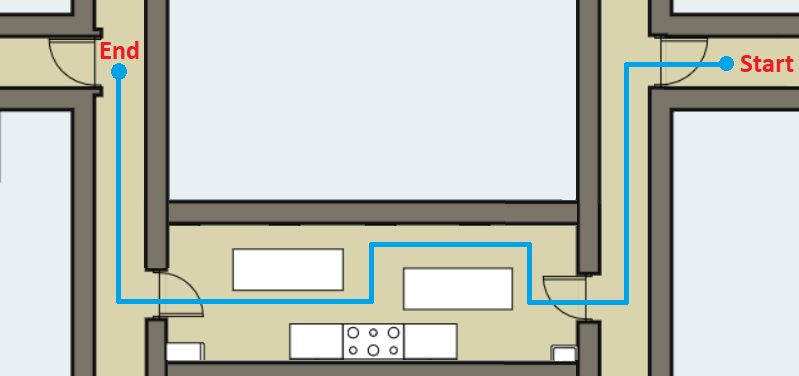}}
\caption{Sketch of the testing route}
\label{fig:parkour4}
\end{figure}

In the second experiment to evaluate the combined system, a testing route was designed. It includes walkways, doors, and obstacles that need to be detected to complete the run. A sketch of the route can be seen in figure \ref{fig:parkour4}. Only the indoor mode of the downsampling algorithm was used and the system was tested in complete darkness. This should showcase the performance of the system in a setting where an optical system would not be able to function. The time required to complete the testing route was recorded for five volunteers. Again three iterations with a week of delay were recorded, to assess the relative improvement over time. The metric is an indicator of the adaption to the system during everyday usage.

\section{Results}
\begin{figure}[htbp]
\centerline{\includegraphics[width=0.5\textwidth]{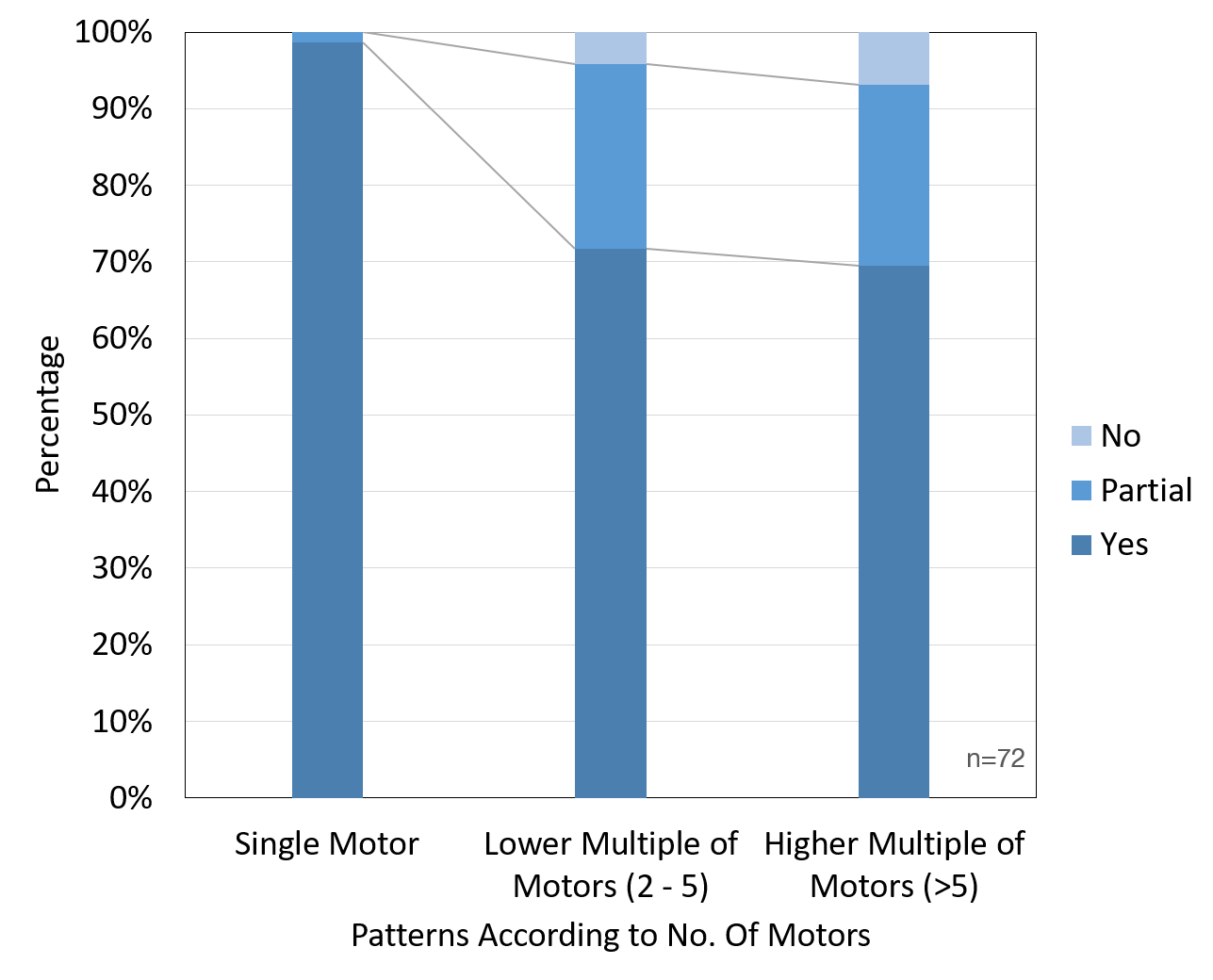}}
\caption{The percentage of correctly identified and localized vibration patterns divided into single motor vibration patterns, vibration patterns with 2-5 motors vibrating at the same time and patterns with more than 5 motors vibrating at the same time.}
\label{fig:patterns1}
\end{figure}

\begin{figure}[htbp]
\centerline{\includegraphics[width=0.5\textwidth]{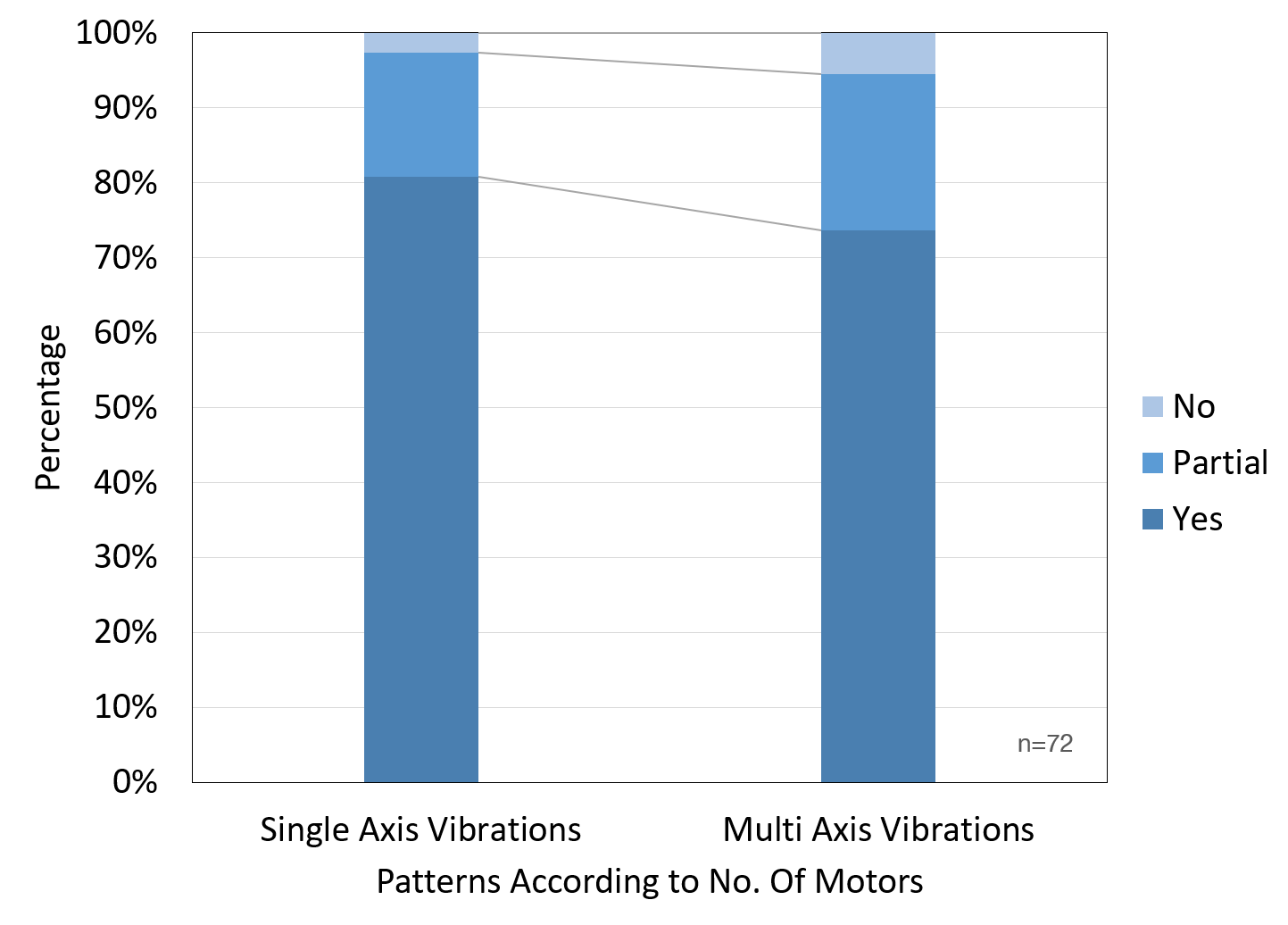}}
\caption{The percentage of correctly identified and localized vibration patterns divided into vibration moving along a single axis and vibration moving along different directions at the same time (multi-axis).}
\label{fig:patterns3}
\end{figure}

The first experiment of evaluating the haptic feedback sleeve showed promising results. Users were able to identify patterns involving single motors with near-perfect accuracy of 98,6\% (figure \ref{fig:patterns1}). The accuracy dropped to $\sim$70\% for patterns with multiple motors vibrating at the same time. The results are similar for lower numbers and higher numbers of multiple vibrating motors. Overall the percentage of incorrectly identifies patterns triples moving from single to lower multiple to higher multiple. The difference can be attributed to the interference when multiple motors vibrate, making it difficult for the user to accurately identify the number of motors. The higher the number of motors vibrating, the higher the interference and hence the higher the number of incorrectly identified patterns. Moreover, as shown in the second graph in figure \ref{fig:patterns3} users performed better on patterns with vibrations along a single axis compared to those along multiple axes. This is probably due to the interference of motors and the difference in sensitivity on various parts of the forearm. Therefore vibrations in the more sensitive parts overshadow vibrations in less sensitive areas.\\

Qualitative user interviews after each test run showed few additional observations: 1) Vibrations over more muscular regions of the forearm are easier to recogniz compared to other regions. This was already anticipated during the experiment setup and therefore users were instructed to wear the band on the inside of the forearm, covering a larger sensitive area. 2) Vibrations close to the wrist are best recognized, indicating high vibrational sensitivity. 3) Most users reported a too small motor distance, making it harder to distinguish motor vibrations. 4) The already good performance in identifying patterns showed no further increase over the weekly iterations. Therefore, the adaption time for the haptic feedback sleeve seems to be very short.

\begin{figure}
\def\svgwidth{\linewidth}
\centerline{\scalebox{.7}{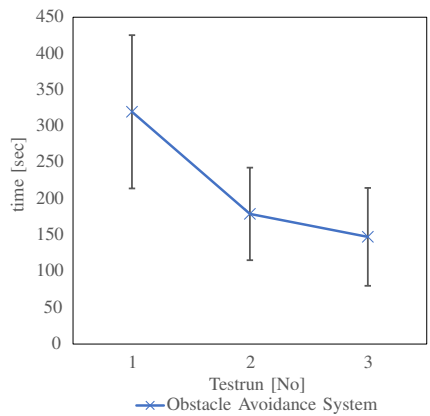}}
\caption{The average time needed to complete the test route for three test-runs (bold color) and the individual test-runs of each user (light color).}
\label{fig:testrun4}
\end{figure}

The results for the second experiment testing the obstacle avoidance system also showed great results. Even in their first run, all five participants were able to complete the obstacle route. In the first run, the total time for completion ranged between 175 and 466 seconds (table \ref{tab2}). Besides person2, who lost orientation for a short time in his last run, all participants show a steady improvement in time over the three runs. The variance in performance between the contestants is quite high, which can be seen in figure \ref{fig:testrun4}. Overall the averages for the runs decreased from 320 seconds in the first run to 148 seconds in the third run, two weeks later. This is a reduction of 53 percent and shows that an obstacle avoidance system would need some time for the user to adapt to the handling. This might be attributed to the highly unusual method of presenting distance information via vibration on the forearm. Since the learning curve in our experiments has not reached saturation, it would also be possible that user performance would increase even more over time.

\begin{table}[htbp]
\caption{Obstacle Avoidance System}
\begin{center}
\begin{tabular}{|c|c|c|c|}
\hline
&Run 1&Run 2&Run 3\\
\hline
person1 [sec] &337 &206 &155\\
\hline
person2 [sec] &340 &229 &238\\
\hline
person3 [sec] &466 &120 &111\\
\hline
person4 [sec] &281 &239 &175\\
\hline
person5 [sec] &175 &102 &59\\
\hline\hline
average time [sec] &320 &179 &148\\
\cline{1-4} 
average time [\%] &100 &60 &47\\
\hline
\end{tabular}
\label{tab2}
\end{center}
\end{table}

\section{Conclusion}
In this paper, we presented a new depth-based obstacle avoidance system. Our solution is low-cost, lightweight and designed for indoor as well as outdoor usage. The haptic feedback sleeve does not interfere with the user's sense of hearing which is extensively used by the visually impaired. It also takes away stigmatization, allowing blind people to add a useful tool for more independence, safety, and confidence. The system could also be a good add-on for usage outdoors, allowing the identification of objects not in the immediate vicinity, which would not be possible with a cane.
The performance tests showed that users are able to identify vibration patterns on their forearms. For single motor vibration patterns, users show an identification accuracy of 98,6\% and even for complex multi-motor and multi-axis vibration the identification accuracy is still 70\% or larger. During the tests, it was discovered that the ability of tell single motor vibrations apart is dependent on the specific location and muscularity of the forearm region. This could be accounted for in further improvements. In the darkness indoor testing route, all participants were able to complete the navigation task. Users improved their time of completion by 53\% over three runs, with one week separating every run. This shows that there is a substantial adaption time necessary for the usage of the system.

Further developments could include the implementation of an object recognition, based on the RGB image of the camera. This would allow the system to artificially generate bordered paths, solving the problem of orientation in a large open space. The proposed system is also already prepared for the implementation of bone conduction speakers, allowing the implementation of voice-based control. This would potentially add a navigation system to the proposed obstacle avoidance. Overall the system is a promising approach towards providing more independence for the visually impaired.



\end{document}